# Spatiotemporal control of surface plasmon polariton wave packets with nanocavities

Naoki Ichiji,[1] Yuka Otake,[1] Atsushi Kubo[2]
[1]*Graduate School of Pure and Applied Sciences,*
*University of Tsukuba,1-1-1 Tennodai, Tsukuba-shi, Ibaraki 305-8573, Japan*
[2]*Division of Physics, Faculty of Pure and Applied Sciences,*
*University of Tsukuba,1-1-1 Tennodai, Tsukuba-shi, Ibaraki 305-8573, Japan*
(Date: March 2021)

**Modulation of the optical index by means of atomic and material resonances provides a basis for controlling light propagation in natural and artificially fabricated materials. In addition, recent advances in the tuning of spatiotemporal couplings of ultrashort laser pulses have enabled almost arbitrary control over the group velocity of light. Here, using femtosecond time-resolved microscopy and numerical calculations, we investigate the spatiotemporal dynamics of a surface plasmon polariton wave packet (SPP WP) that interacts with a plasmonic nanocavity. The nanocavity consists of metal-insulator-metal multilayer films that function as subwavelength meta-atom possessing tunable discretized eigenmodes. When a chirp-induced femtosecond SPP WP is incident on a nanocavity, only the spectral component matching the resonance energy is transmitted. This spectral clipping effect is accompanied by a spatial shift of the WP. The shift can be adjusted in either the positive or negative direction by controlling the resonance energy or the chirp. If this spatial shift is regarded as a modulation of the apparent group velocity in the nanocavity, the range of modulation includes superluminal, subluminal, and negative group velocities.**

Controlling the spatiotemporal dynamics of light pulses has been a fascinating subject in optical physics. The tunability of material dispersions and/or optical resonances has enabled control over group velocities of light, referred to as superluminal, subluminal, and backward light [1], in a variety of materials and artificially fabricated nanostructures, including gaseous atoms [2, 3], ultracold atoms [4], optical fibers [5], ring resonators [6], photonic crystals [7], plasmonic Bragg gratings [8], and metamaterials [9-11]. Metamaterials consist of arrays of subwavelength-scaled optical resonators, each of which provides specific shifts in the phase, amplitude, and polarization between the incident and scattered components of the light field [12]. Accordingly, tailoring light has been realized in many aspects, such as anomalous refraction and reflection [13, 14], vector beam formation [14], spectral filtering [15], light acceleration [16], and ultrafast optical pulse shaping [17]. Within the framework of classical electrodynamics, the refractive index of materials varying near the resonance lines of atoms induces abnormal light propagation in natural materials [2, 18, 19]. In metamaterials, meta-atoms are responsible for the resonant processes, each of which accumulates over the spectral bandwidth and the sectional area of the incident light beam to determine the entire spatiotemporal behavior of the light pulses. In addition, recent advances in controlling the ultrashort light pulses through spatiotemporal couplings [20-22], such as manipulation of the wavevector-frequency relations of structured lights [23-25], tuning the temporal chirp and longitudinal chromatism [26-28], and complex reconstruction of pulse-fronts [29, 30], have enabled almost arbitrary control over the group velocity or peak-intensity velocity of light [31]. Therefore, understanding the dynamic responses of individual meta-atoms and their roles in the modulation of light pulses in terms of the intensity, spatial shape, and temporal shape is indispensable for emerging applications based on the newly developed controllability of light [32, 33].

In this paper, we investigate the femtosecond spatiotemporal dynamics of surface plasmon polariton (SPP) wave packets (WPs) interacting with a subwavelength-scale nanocavity (NC) by using a femtosecond time-resolved two-photon fluorescence microscopy method. Irradiation with 10 fs light pulses initiates SPP WP propagation on a gold (Au) film equipped with a metal-insulator-metal nanocavity (MIM-NC), a typical meta-atom function for visible to infrared light regions [13, 34-37]. Conversion of light pulses to SPP WPs confines light fields to a two-dimensional surface [38] and enables visualization of electromagnetic waves by transforming a portion of the evanescent fields into emissions [39]. Time-resolved movies reveal the motion and deformation of SPP WPs evolving as a function of the pump-probe delay time. Particularly, upon transmission through the MIM-NC, the coordinates of the intensity peak of a WP showed a distinct spatial shift relative to the coordinates of a reference peak. The spatial shifts were controlled to within several micrometers in either the forward or reverse directions depending on the cavity eigenenergies tuned by the structural length of the MIM-NC. Simulations using a finite-difference time-domain (FDTD) method and an analytical model based on a complex dispersion (CD) relation of SPPs (CD model) revealed that spectral clipping for a chirp-induced SPP WP provoked spatial shifts, which led to large changes in the effective peak velocity of the WP in the MIM-NC. Because MIN-NCs transmit only the spectral components falling within the resonance line width, they effectively block SPP waves when the eigenenergies are detuned from the spectral range of the SPP.

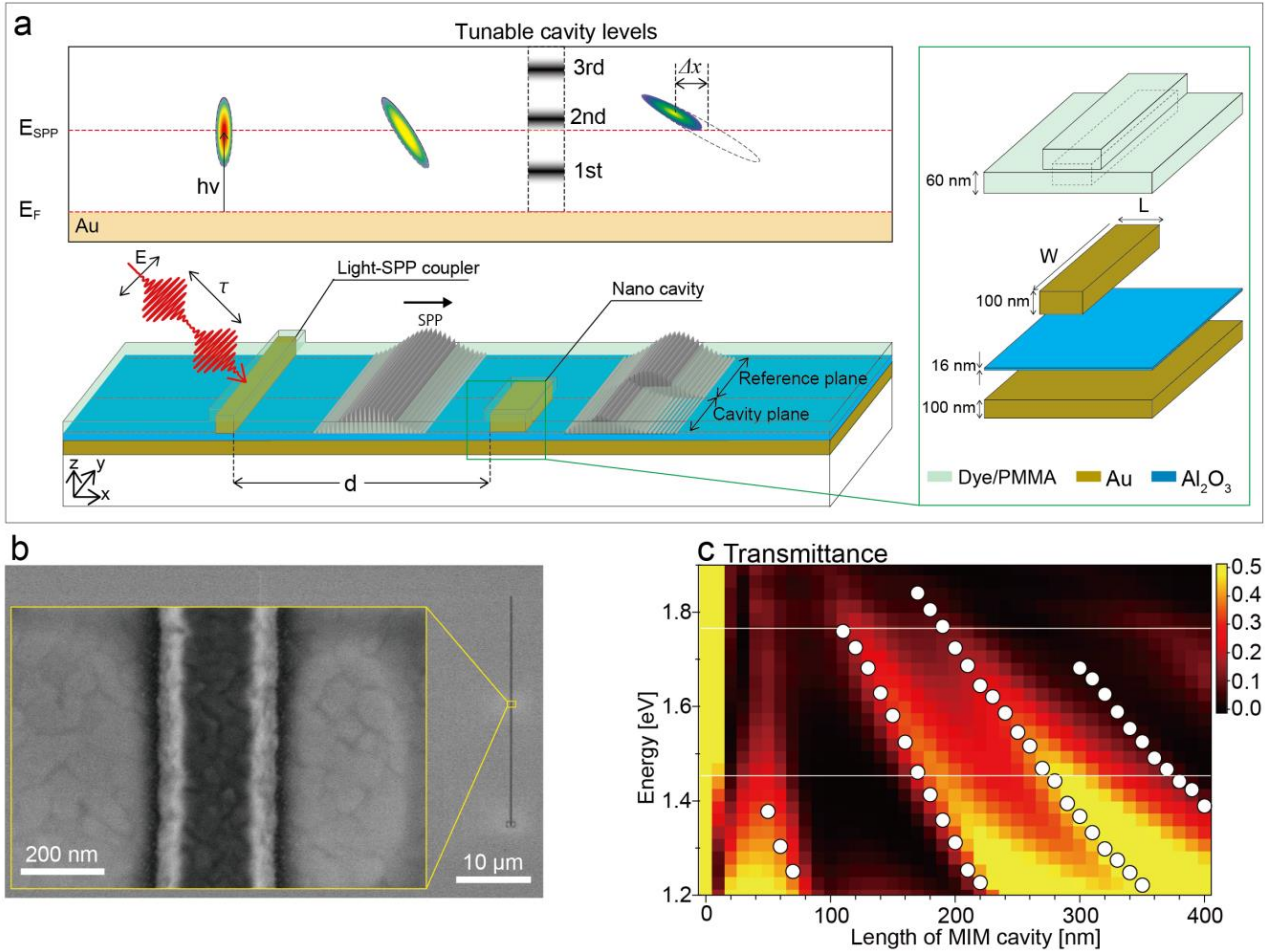

**FIG. 1 Excitation of an SPP WP on a Au surface with an MIM-NC.** (a) Schematic of the sample equipped with an MIM nanocavity and a light-SPP coupler. The sample surface is covered with a dye-doped PMMA thin film. Irradiation with the 10-femtosecond laser light excites the SPP WP at the energy level $E_{SPP}$, which is comparable to the light energy, $h\nu$ (inset). The SPP WP has a considerably broad energy distribution reflecting the spectral bandwidth of the laser. Resonant interaction with an MIM-NC, which possesses a few discretized energy levels near $E_{SPP}$, results in a spectrally clipped transmission, leading to a spatial shift ($\Delta x$) in the position of the intensity peak of the WP. (b) SEM image of the MIM-NC with a length $L$ = 220 nm. (c) Two-dimensional plot of transmittance spectra of SPP for the MIM-NC with a length $L$ = 10~220 nm plotted as a function of $L$ and energy (Supplementary Fig. S1). The four series of solid circles indicate peak energy of the 1st ~ 4th order eigenmodes of the MIM-NC. The two white lines indicate the upper and lower spectral limits of the excitation laser.

**Results**

**SPP WPs and MIN-NCs.** The MIM-NCs possessing a few discretized eigenmodes in the spectral range of the excitation pulse (white solid lines in Fig. 1(c)) were prepared by placing rectangular Au nanoblocks on a sapphire (thickness, $h$=16 nm)/Au ($h$=100 nm) film (Fig. 1(a)). The levels of the eigenmodes were tuned by adjusting the length ($L$) of the cavity in the range of 50~220 nm. SPP WPs were launched from a light-SPP coupler by irradiating with P-polarized, dispersion compensated 10 fs, 810 nm pump-probe pulsed light at an incident angle ($\theta_i$) of 45° from the surface normal. Here, we focus on the dynamics of the pump-excited SPP WP propagating along the positive $x$-axis. The probe pulse, temporally separated from the pump pulse by a delay time ($\tau_d$), interferes with the SPP WP on the surface to form a transient grating of the surface field when they overlap both spatially and temporally [40, 41]. The spatial map of the grating, which contains the dynamical information of the SPP WPs, was converted to the radiation field via the two-photon fluorescence of the dye-doped poly(methyl methacrylate) (PMMA) layer and then detected by an optical microscope equipped with a charge-coupled device (CCD) camera [39]. The width ($W$) of MIM-NCs was chosen so that half the length of the wavefront of the SPP WP interacts with the MIM-NC ("cavity plane" in Fig. 1(a)), while the other half continues to propagate on the flat surface ("reference plane") for comparison. This system can be regarded as an optical cavity coupled with a waveguide [42, 43]: The incident SPP WP couples to the cavity mode and is subsequently emitted from the other side if the frequency component coincides with the eigenenergy of the cavity (Fig. 1(a) inset) [44]. The transmittance spectra of SPPs through an MIM-NC with a length $L$ = 10 ~ 400 nm were evaluated by the FDTD method (Fig. 1(c) and Supplementary Fig. S1) [45]. The spectral shapes are characterized by a Fano-type asymmetric line shape [46-48] with transmittance maxima near the eigenenergies (open circles in Fig. 1(c)). The transmittance is significantly depressed to approximately 0.1 for $L$ ~100 nm, corresponding to a valley between the first- and second-order resonances and suggesting that the MIM-NC functions as a subwavelength reflector. For longer $L$ values, color-selective transmissions are expected.

**Time-resolved imaging of SPP WPs.** Figure 2 shows images of time-resolved micrographs obtained for MIM-NC lengths of (a), 140 nm, (b), 160 nm, and (c), 220 nm at a pump-probe delay $\tau_d$ = 176 fs, at which SPP WPs passed through the MIM-NCs and reached $x$ = 60 μm. Cross-sections of beat patterns in Figs. 2(a), 2(b), and 2(c), prepared by addition-averaging the images along the $y$-direction over the areas indicated by two blue solid lines (cavity planes), are shown in Figs. 2(d), 2(e), and 2(f). For a controlled comparison, cross-sections were also taken in the flat regions indicated by red solid lines (reference planes) in a similar manner.

While the beat patterns at the reference planes showed almost identical features for the $L$ = 140, 160, and 220 nm cavities, those at the cavity planes varied, and the variations were very sensitive to $L$. The intensity of the beat pattern largely decreased (~0.2) at $L$ = 140 nm but remained considerable (>0.5) at $L$ = 160 nm and 220 nm. The WPs were also associated with changes in envelope shapes. For the 220 nm cavity, the beat pattern at the cavity plane was well fitted with a single Gaussian-modulated sinusoidal wave function with a peak position and width of the envelope comparable to those for the reference plane. A single sine-Gaussian wave shape was also found for the 160 nm cavity; however, the envelope peak showed a distinct negative shift with respect to the reference. For the 140 nm cavity, the wave was reasonably fitted by a superposition of two sine-Gaussian functions with the envelope peaks shifted by a 5 μm advancement and a 7 μm retardation with respect to those of the reference.

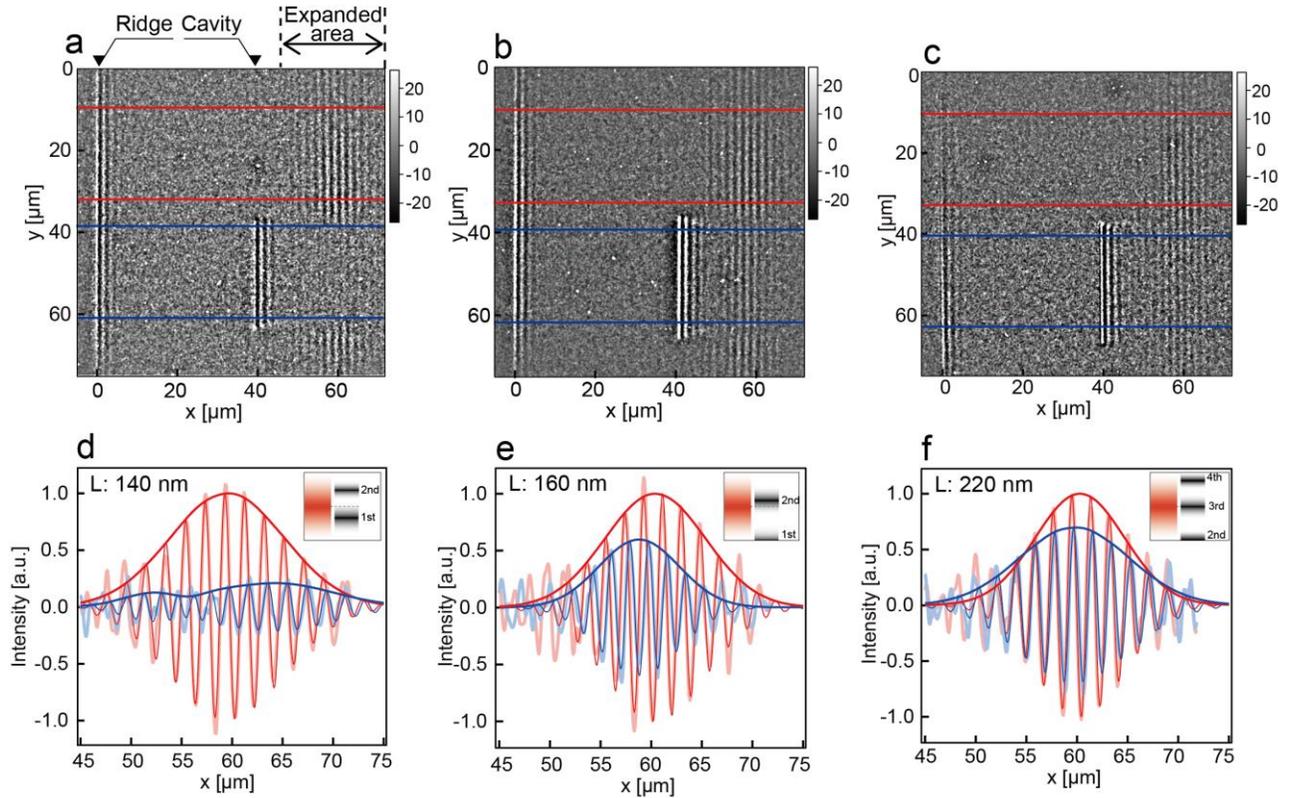

**FIG. 2 Fluorescence images of transmitted SPP WPs.** Images of time-resolved two-photon fluorescence microscopy of SPP WPs which transmitted through an MIM NCs with the length $L$ = 140 nm (a), 160 nm (b), and 220 nm (c), respectively. The beat patterns constructed by interference between a SPP WP and a probe pulse, which reveal the spatial distributions of the SPP WP, were extracted by removing background and delay time independent intensities (Supplementary Fig. S5). Red and blue lines indicate the reference and cavity planes, respectively. Cross sections of beat patterns at the reference and cavity planes in (a), (b), and (c) were prepared by addition averaging the images along $y$-direction and shown by the thick light red and light blue lines in (d), (e), and (f), respectively. Their least-square fits and the envelopes are shown by thin dark red and dark blue lines. The cross sections in all reference planes ($L$ = 140, 160, and 220 nm) and those in cavity planes with $L$ = 160 and 220 nm were fitted by single Gaussian-sine function, but that in the cavity plane with $L$ = 140 nm was fitted by a sum of two Gaussian-sine functions. The energy distributions of a spectrum of SPP WP and eigenmodes of a MIM-NC are schematically shown in insets of (d-f). The MIM-NCs with $L$ = 140, 160, and 220 nm provide interactions specified as; a valley of two resonance modes (d), a slightly detuned resonance to higher energy side (e), and an almost on-resonance (f), respectively. (Full frames of a time-resolved movie for $L$ = 180 nm MIM-NC is shown in Supplementary Movie1.)

The spatial shifts in the intensity maxima of SPP WPs caused by MIM-NCs were further investigated by examining a sequence of time-resolved micrographs (Supplementary Video 1). Figures 3(a) and 3(b), respectively, show time-resolved series of cross-sections of the beat patterns taken from the reference and the cavity planes for an MIM-NC with $L$=160 nm. The spatial shape of a beat pattern essentially reflects that of an SPP WP itself (Supplementary Figs. S4 and S5). While the beat pattern in the reference area (Fig. 3(a)) showed a propagation of the intensity peak (yellow circles) at a constant rate, that in the cavity area (Fig. 3(b)) revealed a distinct peak shift compared to the reference (green dotted line) after transmission through the MIM-NC (*i.e.*, after $x = 40$ μm). The coordinates of the intensity maxima of the beat patterns are plotted in Fig. 3(c) for $\tau_d = 85.7 \sim 214.7$ fs. The series of cavity planes showed a homogeneous negative shift of -1.83 μm after passing through the cavity (Fig. 3(d)). The group velocities of the SPP WPs were evaluated as $v_g^{SPP}$=1.75×10$^8$ m/s (0.58$c$, $c$: speed of light in a vacuum).

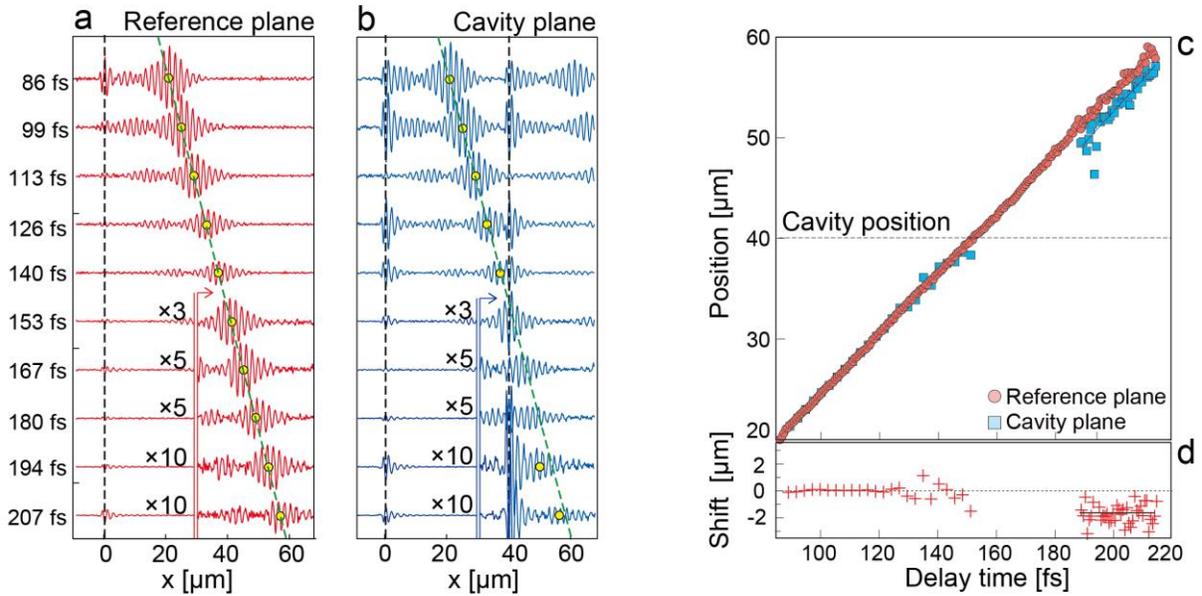

**FIG. 3 Temporal evolution of an SPP WP.** Cross sections of the time-resolved two-photon fluorescence images taken from (a); a reference plane, and (b); a cavity plane, equipped with an MIM-NC with a length $L = 160$ nm are shown. The shape and coordinate of a beat pattern reflects the spatial form of a SPP WP through an interference of the SPP WP and probe pulse. The range of the delay time is $\tau_d = 86\sim207$ fs with an interval of 13.5 fs or $5 \times 2\pi$ rad of the carrier wave of excitation laser. The yellow circles indicate coordinates of the intensity maxima of beat patterns. The green dotted line in (a), indicating a constant advancement of a SPP WP, is duplicated and plotted in (b). (c) Coordinates of the intensity peaks of beat patterns taken from the reference plane (red circles) and cavity plane (blue squares) are plotted as the function of delay time. The intensity maxima in the cavity plane exhibit a homogeneous negative shift compared to those in the reference over a region where the isolated wavepacket forms are recovered ($x \geq 50$ μm). The amount of shifts ($\Delta x$) is shown in (d).

## Discussion

A spatial shift of the SPP WP in the negative direction means that the SPP WP was delayed in time or that the apparent group velocity decreased. The systematic survey at cavity lengths $L = 50 \sim 220$ nm showed that the shift in the transmitted SPP WP can be tuned by a few microns in either the forward (red-colored area) or backward (blue-colored area) direction, providing an apparent superluminal or subluminal SPP depending on $L$ (Fig. 4(e)). The distance of the maximum shift in the forward direction (~3 μm) well exceeded the cavity length ($L =110$ nm), exhibiting phenomenological similarity to a material with a negative group velocity, by which the emergence of an exit pulse before the arrival of an input pulse was realized [1, 2]. Spatial shifts toward the forward and backward directions were observed for cavity lengths $L = 90 \sim 110$ nm (red-colored area in Fig. 4(e)) and $150 \sim 220$ nm (blue-colored area), respectively. At a boundary ($L = 120 \sim 140$ nm, purple-colored area), the amplitudes of transmitted SPP WPs attenuated (Fig. 4(f)), and the envelopes deformed to the point at which they were reasonably fit by the sum of two Gaussian curves.

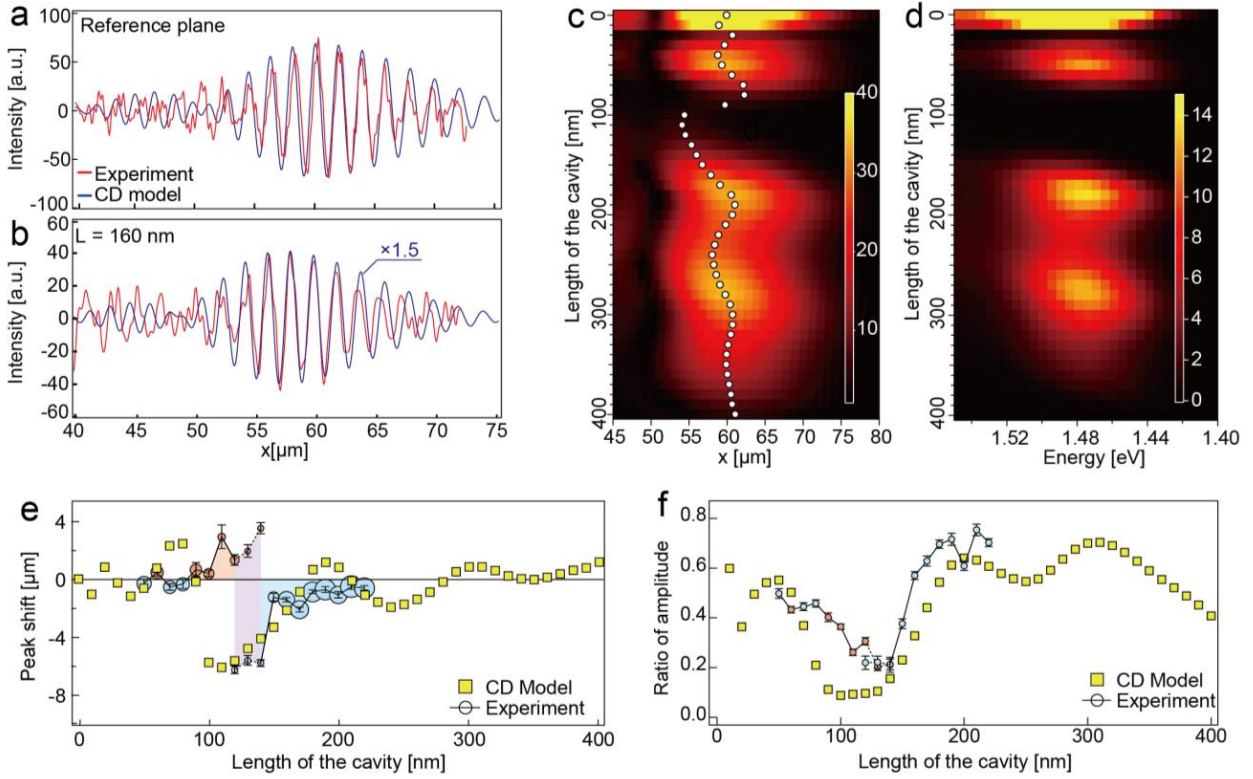

**FIG. 4 Peak shift, spectral modulation, and amplitude attenuation of an SPP WP by CD model and experiment.** Calculated beat patterns using CD model for the reference plane; (a), and for the cavity plane; (b), equipped with an MIN-NC with a length $L = 160$ nm (blue lines), and their corresponding experimental results taken from the time-resolved images (red lines) are shown. The delay time ($\tau_d$) is 185 fs, at which the intensity peak of a beat pattern in the reference plane reaches to $x = 60$ μm for both the calculations and experiments. (c) Two-dimensional plots of the envelope of a beat pattern calculated by the CD model as a function of the length of the cavity ($L$) and $x$. The white circles indicate the peak positions of each envelope. (d) Two-dimensional plots of the spectra of transmitted SPP WPs calculated by the CD model as a function of $L$ and energy. (e) The extent of the peak shift of the beat pattern in the cavity plane compared with that in the reference plane. The yellow squares and colored circles show the peak shifts determined by the CD model and the experiment, respectively. The red and blue circles indicate the forward and rearward peak shifts, respectively. The diameter of a circle indicates an amplitude of a beat pattern. For the length of cavity $L = 120 \sim 140$ nm (purple colored region), peak shifts are indicated by two circles reflecting the double-peak feature (Fig. 2d) of an experimental beat pattern. (f) The ratio of the maximum amplitude of a beat pattern in the cavity plane to that in the reference plane plotted as a function of $L$ The yellow squares and colored circles show results of the CD model and the experiment, respectively. (Supplementary Figs. 3 and 4 and Video 2)

We attribute the spatial shifts to spatiotemporal clipping of a chirped SPP WP enforced by the MIM-NC [45, 49]. Because of the normally dispersed character of the SPP mode, the instantaneous carrier frequency of an SPP WP gradually becomes up-chirped during propagation on the Au surface. While the intensity peak continuously propagates at a group velocity defined at the center frequency, the wave components of higher (lower) frequencies proceed at slower (faster) group velocities. A temporal chirp in frequency is correlated with a spatial chirp in wavenumber. When an SPP WP passes an MIM-NC that possesses a resonance frequency coinciding with the higher (lower) part of the spectral bandwidth, the peak position of the transmitted WP shifts backward (forward). If two different resonances contribute, then the WP is expected to have two separate peaks, as seen in Fig. 2(d).

To obtain more detailed insights into the deformation of the WPs, we performed model calculations based on CD relations of SPP modes (CD model, Supplementary Figs. S3 and S4, and Video 2) [40, 50]. The temporal waveform of an SPP WP at position $x$, $E_{SPP}(x,t)$, was retrieved from the Fourier expansion coefficients, of which the amplitudes, $R(x,\omega_i)$, and phases, $\phi(x,\omega_i)$, were evaluated by using the CD relation of the SPP as $R(x,\omega_i) = R(0,\omega_i) \cdot \exp(-k''_{SPP}(\omega_i) \cdot x)$ and $\phi(x,\omega_i) = \phi(0,\omega_i) + k'_{SPP}(\omega_i) \cdot x$, respectively. $k'_{SPP}$ and $k''_{SPP}$ are the real and imaginary parts of the complex wavevector of the SPP, respectively. The resonant interactions of SPP WPs with MIM-NCs were modeled by taking into account the amplitude attenuations according to the transmission spectra of the MIM-NCs and the phase accumulations thorough the Lorentzian resonances and wave propagation in the cavity. In our experiment, the fluorescence from the dye-doped PMMA layer was emitted through two-photon excitations. Therefore, we attributed the intensity of the fluorescence to a time integral of the fourth power of the total electric field: $I_{Calc}(x) = \int_{-\infty}^{\infty} \big(|E_{SPP}(x,t) + E_{Light}(x,t)|^2\big)^2 dt$, where $E_{Light}$ is the surface field directly induced by the excitation laser.

Figures 4 (a) and 4(b) show simulated beat patterns superimposed on experimental patterns obtained from a reference plane and from a cavity plane with a 160 nm cavity, respectively. The amplitude of the simulated result in Fig. 4(b) was multiplied by a factor of 1.5 to reflect the distribution of an oval-shaped light field on the sample surface. The agreement between the CD models and the experiments was good. In particular, a backward shift in a peak position associated with an asymmetric spatial distribution of the envelope shape of the transmitted WP was reasonably reproduced. The envelope shapes of the beat patterns calculated for the range $L = 0 \sim 400$ nm are shown in Fig. 4(c). For the range $L \geq 20$ nm, the plot of envelope shapes contains three elliptically shaped strength distributions, which correspond to the 1st – 3rd resonance modes of the MIM-NCs. Each distribution was slanted off the horizontal axis: The peak position (open circles in Fig. 4(c)) shifted backwards as the cavity length decreased and then gradually shifted forward as $L$ increased.

The relation of the peak position to $L$ correlated closely with that of the peak energy to $L$. Figure 4(d) shows the spectra of SPP WPs transmitted through MIM-NCs with $L = 0 \sim 400$ nm obtained from Fourier transforms of temporal waveforms of WPs at $x = 50$ μm. Overall, the configuration of Fig. 4(d) is very similar to that of Fig. 4(c). The peak energy is blueshifted (red) for an SPP WP with a retarded (forwarded) peak position, consistent with the mechanism of spatiotemporal clipping of an up-chirped SPP WP by an MIM-NC. As shown in Fig. 4(e) and 4(f), the amount of the peak shift of the SPP WP with respect to the reference and the relative amplitude calculated by the CD model were reasonably consistent with the experimental results.

Controlling the spatial coordinates of the transmitted SPP WPs can be regarded as an application of spatiotemporal couplings (STCs) [20, 21]. In the abovementioned case, the STC is simply an up-chirping of the carrier frequency of an SPP WP that possesses longer (shorter) wavelengths in the front (tail) region. The intensity peaks of SPP WPs were controlled by tuning the eigenenergy of the MIM-NC. In the following, we discuss another strategy for controlling the coordinates of intensity peaks by using the tunability of STCs with external chirp control of the excitation light pulses. We calculate the shifts in the transmitted SPP WPs using the CD model by applying under- or overcompensated chirp conditions to the excitation source. The phase distribution of a chirped pulse at $x=0$, $\phi_c(0,\omega_i)$, is given as $\phi_c(0,\omega_i) = \phi(0,\omega_i) + \phi_g(\omega_i)$, where $\phi_g(\omega_i)$ is the additional phase for the chirped pulse. Light passing through a dispersive material is given by $\phi_g(\omega_i) = (\omega_i/c) \cdot n_g(\omega_i)g$, where $n_g$ is the refractive index and $g$ is the thickness of the material. We assumed the material to be fused silica glass. The overcompensation conditions were created by simply adapting negative $g$ values. The MIM-NC was assumed to have a length $L = 140$ nm with a 2nd-order resonance at 1.7 eV. This cavity mode provides a transmission window to a higher-energy region of a spectrum of a 10 fs, 810 nm Gaussian-shaped excitation pulse. Figure 5 (a) shows the waveforms of the chirped excitation pulses for several glass thicknesses and the corresponding envelope shapes of SPP WPs at 500 fs after the excitations. For a glass thickness of $g = 0$ mm, the intensity peak was retarded compared to the reference, as expected for an up-chirped SPP WP. The extent of the retardation shift increased as $g$ increased but decreased to ~0 for a negative glass thickness of $g = -10$ mm, corresponding to a group velocity dispersion (GVD) of -360 fs$^2$. This null shift occurred because the provided negative GVD compensated for the positive dispersion of the SPP mode on the Au surface. For more negative $g$ values, the SPP WPs maintained down-chirps when they reached the MIM-NC, resulting in forwarding of the intensity peaks. The shifts in the intensity peaks calculated for $g = -20$~$10$ mm (GVD: -720~360 fs$^2$) are plotted in Fig. 5 (b). Continuous control over the spatial shift of the transmitted SPP WP was achieved in the range of -8~5 μm. Notably, the transmitted WPs maintained envelope shapes and pulse widths comparable to those of the reference WPs, despite the spectral clippings by the MIM-NC inducing a modulation in their center frequency. Therefore, if one ignores the instantaneous frequency and measures only the temporal evolution of the intensity of a transmitted SPP WP by a point detector, the induced shift will be recognized as a change in the transit time or the apparent group velocity ($v_g^{ap}$) of the WP in the MIM-NC. Using the spatial shift, $\Delta x$, and the group velocity of SPPs on the flat surface, $v_g^{SPP}$, the apparent group velocity in a cavity is $v_g^{ap} = v_g^{SPP} \cdot L/(L - \Delta x)$ [2]. The ratio of $v_g^{ap}$ to $v_g^{SPP}$, which will give the apparent group-velocity index, $n_g^{ap} = v_g^{ap}/v_g^{SPP}$, changed from -60 to 90 as $g$ changed, suggesting that manipulating STCs using external chirp controls provides a wide-ranging tunability of the apparent group velocity of SPP WPs, including the superluminal, subluminal, and negative group velocity regions.

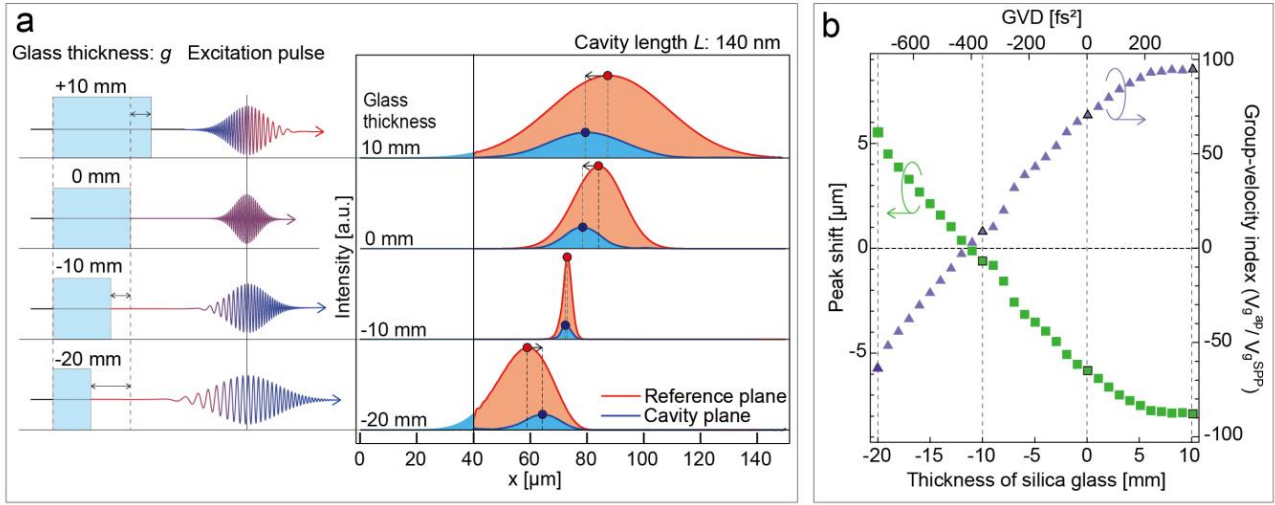

**FIG. 5 Control of spatial shift of an SPP WP by frequency chirp of the excitation pulse.** (a) Envelope shape of an SPP WP calculated by the CD model for a reference plane (red graph) and that for a cavity plane (blue graph) equipped with a MIM-NC with a length $L$ = 140 nm are shown for several chirped pulse conditions. Red and blue circles indicate coordinates of the intensity maxima of SPP WPs in the reference plane and the cavity plane, respectively. Temporal waveform of a chirp-induced excitation pulse prepared by adding a group velocity dispersion (GVD) which amounts to an effective thickness of a silica glass $g = -20$~$10$ mm are schematically shown. (b) The extent of the peak shift of a SPP WP after the transmission of the MIM-NC and the apparent group velocity index of SPP WP in the MIM-NC are shown as a function of the GVD or the thickness of a silica glass applied to the excitation pulse.

**Conclusion**

In conclusion, spatiotemporal dynamics of SPP WP interacting with MIM-NC possessing discretized eigenenergies near a spectral distribution of WP was investigated using time-resolved microscopy and numerical calculations. Spatial distributions of the surface electric fields visualized with femtosecond temporal and micrometer spatial resolution via two-photon fluorescence emission revealed the distinct spatial shifts in the peak position of WP after passing through a MIM-NC. Analyses of the transmittance spectra of a SPP WP through a MIM-NC using FDTD simulations and calculations of the waveforms of SPP WP using the CD model showed that the peak shifts were induced by spatiotemporal clipping of a chirped SPP WP enforced by the MIM-NC. The peak shift can be controlled to within several micrometers in either the positive or negative directions by adjusting the eigenenergy of a MIM-NC or the chirp of an excitation pulse, that phenomenologically comparable to a tunability of the apparent group velocity index of SPP over a range of plus or minus a few tens. The controllability of the SPP WP shown here is an application of the spatiotemporal couplings and will provide useful insights into the developments of novel optical devices based on the complex control of light waves such as structured lights.

**Methods**

**Sample fabrication**

The sample was prepared in the following manner. First, a Au layer with a thickness of 100 nm and a subsequent sapphire ($Al_2O_3$) layer with a thickness of 16 nm were deposited on a silicon substrate by sputtering and atomic layer deposition. Next, rectangular Au blocks with a width ($W$) of 30 μm, a thickness ($h$) of 100 nm, and a length ($L$) ranging from 50 to 220 nm were placed by using standard electron beam lithography and thermal evaporation methods. To visualize the spatial intensity profiles of the surface electromagnetic fields, the whole area of the sample was coated with dye (coumarin 343)-doped PMMA by a standard spin coating technique. The thickness of the PMMA layer was determined by a spectroscopic ellipsometer. To prevent unintentional loss of the fluorescence intensity due to bleaching of dye molecules, the PMMA layer was occasionally re-coated throughout the experiment. The thickness was 50 nm for the experiments shown in Figs.2 and 4 and Supplementary Movie 1, and 60 nm for Fig.3. The uncertainty of the thickness was 5 nm including the ununiformity of the film. The difference in film thickness of 5 nm results in a difference in the group velocity of SPP about $0.1\times10^8$ m/s for the experimental condition used here.

**Femtosecond laser-excited time-resolved two-photon fluorescence microscopy**

We used 10 fs light pulses to excite SPP WPs and to visualize their spatial maps through the light-SPP interferences. Light pulses from a home-built Ti:sapphire laser oscillator (duration, 10 fs; carrier wavelength, 810 nm; repetition rate, 90 MHz; average power, 450 mW) were formed to coaxially align pump-probe pulse pairs by using a Mach-Zehnder interferometer [26]. Then, the pulses were loosely focused to an oval-shaped, 80 μm×60 μm spot on the sample surface, where both an MIM-NC and a light-SPP coupler, a one-dimensional Au ridge separated from the MIM-NC by 40 μm, were placed. P-polarized, dispersion-compensated light pulses were directed to the sample with an incident angle ($\theta_i$) of 45° from the surface normal. The spatial map of the two-photon fluorescence from the dye-doped PMMA layer was detected by an optical microscope equipped with a CCD camera.

**Acknowledgements**

This work was supported by NIMS Nanofabrication Platform in Nanotechnology Platform Project, (JPMXP09F-17-NM-0068), the JSPS KAKENHI (JP14459290, JP16823280, JP18967972, JP20J21825), JST CREST (JPMJCR14F1), and MEXT Q-LEAP ATTO, Japan.
The authors thank H. T. Miyazaki and T. Kasaya for valuable discussions and notable contribution for sample fabrications.